\documentclass[12pt]{article}
\usepackage{latexsym}
\title{Is zero-point energy physical? A toy model for Casimir-like effect}
\author{Hrvoje Nikoli\'c \\
Theoretical Physics Division, Rudjer Bo\v{s}kovi\'{c} Institute, \\
P.O.B. 180, HR-10002 Zagreb, Croatia \\
{\normalsize e-mail: hnikolic@irb.hr} \\
\makebox[1in]{} \\
}
\date{\today}
\begin{document}
\maketitle
\begin{abstract}
Zero-point energy is generally known to be unphysical. 
Casimir effect, however, is often presented as a counterexample, giving rise to a conceptual
confusion. To resolve the confusion we study foundational aspects of Casimir effect
at a qualitative level, but also at a quantitative level 
within a simple toy model with only 3 degrees of freedom. 
In particular, we point out that Casimir vacuum is not a state without photons, and not a ground state for a Hamiltonian
that can describe Casimir force. Instead, Casimir vacuum can be related to the photon vacuum
by a non-trivial Bogoliubov transformation, and it is a ground state only for an effective 
Hamiltonian describing Casimir plates at a fixed distance. At the fundamental microscopic level,
Casimir force is best viewed as a manifestation of van der Waals forces.
\end{abstract}

\vspace*{0.5cm}
Keywords: zero-point energy; vacuum; Casimir effect; van der Waals force 

\newpage

\begin{quote}       
{\it So, in the discussion session
after Casimir's lecture I switched topic and asked: ``Is the Casimir effect due to the
quantum fluctuations of the electromagnetic field, or is it due to the van der Waals
forces between the molecules in the two media?'' Casimir's answer began, ``I have
not made up my mind.''} 

(I.H. Brevik, from the Foreword in \cite{buhmann1}.)
\end{quote}

\section{Introduction}

In physics we measure energy differences, not absolute energies. Zero-point energy,
that is energy of the ground state, is therefore unphysical and can be removed by a simple subtraction.
Yet Casimir effect \cite{casimir},
in its simplest form described as an attractive force between electrically neutral plates,
is often presented as a demonstration that 
zero-point energy is physical \cite{itz-zub,BD,milonni,milton,zin-just,mukhanov,zee,schwartz}.
On the other hand, Casimir effect can also be explained as a force that originates from 
van der Waals forces
\cite{casimir-polder,lifshitz,lifshitz2,LL9,mostepanenko-rmp,mostepanenko-advances,buhmann1,buhmann2,commins},
without referring to zero-point energy.
There are arguments that the explanation in terms of van der Waals forces is more fundamental
\cite{jaffe,nik_casimir} (see also \cite{buhmann1,buhmann2}),
yet some consider the question of the true nature of Casimir effect as a 
controversy \cite{lamoreaux-pt,brevik-milton,padmanabhan} that still needs to be resolved.
%
%

A part of the difficulty stems from the fact that explanation of Casimir effect requires 
quantum electrodynamics (QED), which involves various technical difficulties
coming from the fact that QED is a quantum field theory,
i.e. a theory with an infinite number of degrees of freedom.
To overcome this technical difficulty, in this paper we shall not be so much concerned
with technical details of Casimir effect itself.
Instead, our main goal will be to understand in detail how a Casimir-like effect 
can emerge {\em in general}, using only general properties of quantum mechanics.
For that purpose we shall study a toy model with only 3 degrees of freedom,
which under certain approximations can be reduced to 2 or even 1 degree of freedom.
The toy model will be chosen such that it has many conceptual similarities
with the real Casimir effect, but is technically much simpler than that.
This will enable us to understand relatively easily where the effect comes from,
and how is it related to the zero-point energy. 
In addition, to make a contact with actual Casimir physics, 
we shall discuss various aspects of Casimir effect at a qualitative non-technical level. 

The paper is organized as follows. 
In Sec.~\ref{SECmain} we study various conceptual aspects of Casimir effect.
In particular, we explain the difference between ground state and vacuum,
point out that Casimir force can be attributed to the interacting vacuum energy but not 
to the ground state energy, discuss the role of dielectric constant 
and its relation to van der Waals force, and make a motivation for the toy model
discussed in the following sections. 
In Sec.~\ref{SECmodel} we introduce our toy model and analyze its classical properties.
In Sec.~\ref{SECforce} we explain how the toy model leads to a quantum Casimir-like force,
using both a Casimir-like and a Lifshitz-like approach.
In Sec.~\ref{SECcontent} we explore the content of the interacting vacuum
by using the Bogoliubov-transformation method.
In Sec.~\ref{SECreal} we discuss how the calculations in our toy model are related to the calculations
for real Casimir effect.
Finally, the conclusions are drawn in Sec.~\ref{SECconcl}.

As a disclaimer, we also want to remark that in this paper we do not study the relevance of ground-state energy 
to gravitational physics.  
For a possible relevance of Casimir energy to gravitational phenomena 
see e.g. \cite{rovelli} and references therein.

\section{Basic conceptual questions}
\label{SECmain}

One of the main messages of this paper, in agreement with \cite{jaffe,nik_casimir},
is that, at the fundamental microscopic level, Casimir effect should be viewed 
as a manifestation of van der Waals forces, and not as a manifestation of zero-point energy. 
But why then the effect is often attributed to zero-point energy and why such a description works 
fine too? And what exactly are drawbacks of the zero-point energy description?
In this section we give a non-technical answer to those and many other related conceptual questions.

\subsection{What is vacuum?}

In physics, the word ``vacuum'' has many different meanings. 
It can mean a state without any particles whatsoever, or a state without only one kind of particles
such as photons, or a state annihilated by some lowering operators, or a local minimum of a classical potential,
or the state with the lowest possible energy. Of course, all these notions of ``vacuum'' are closely related, 
but the point is that they are not strictly identical. 

Which of those notions of ``vacuum'' is relevant for the description of Casimir effect? Clearly,
Casimir vacuum is not a state without any particles whatsoever, because Casimir effect involves 
plates made of atoms. As we shall see, Casimir vacuum is also not a local minimum of a classical potential,
and perhaps more surprisingly, not even a state without photons. We shall see that Casimir vacuum 
is a state annihilated by some lowering operators which are {\em not photon} lowering operators.

\subsection{Is Casimir vacuum a ground state?}
\label{SECground}

A ground state is a state with the lowest possible energy, and   
energy can be defined as an eigenvalue of a Hamiltonian. 
But {\em which} Hamiltonian? There are many different Hamiltonians used in physics, and hence there are many different 
notions of ``energy''. As a consequence, there are many different notions of ``ground state''.
The notion of a ground state has no meaning at all if one does not specify the Hamiltonian.

Some Hamiltonians are meant to be fundamental, describing ``all'' physics, or at least a large part of physics. 
Other Hamiltonians are merely effective Hamiltonians, describing only a small subset of all physical phenomena. 
Consequently, some ground states are supposed to be fundamental, while other ground states are merely effective ground states.

The Casimir vacuum is one such {\em effective ground state}. It is the lowest energy state 
for the system with Casimir plates at a {\em fixed} distance $y$. 
The existence of the attractive Casimir force implies that plates separated by smaller 
distance $y$ have smaller energy, so energy of plates at a given non-zero distance $y$
cannot be the lowest possible energy. 
Hence, Casimir vacuum is not the ground state for the Hamiltonian describing the change of $y$.
Furthermore, a state without any plates whatsoever has even smaller energy,
so Casimir vacuum is certainly not the ground state for the fundamental Hamiltonian that 
describes all possible physical states, including those with no plates at all.
Casimir vacuum is the ground state for an {\em effective Hamiltonian}, 
a Hamiltonian that describes only those phenomena for which (i) the existence of Casimir plates
is given and (ii) the distance $y$ between the plates is fixed.

\subsection{Can Casimir force be explained in terms of ground-state energy?}

No. By the Newton second law, the force in the $y$-direction creates acceleration $\ddot{y}(t)$.
Hence, to describe Casimir force, $y$ needs to be treated as a dynamical variable, 
not as a fixed parameter. 
Parameter $y$ cannot simultaneously be fixed and non-fixed. 
If it is fixed then Casimir energy can be interpreted as an effective ground-state energy
as explained in Sec.~\ref{SECground}, but in that case there is no Casimir force. 
If it is not fixed then there is Casimir force, but in that case Casimir energy cannot be interpreted as an effective
ground-state energy. 
To make $y$ dynamical, one must construct a new Hamiltonian by adding the appropriate kinetic 
term to the effective Hamiltonian for fixed $y$. 
The Casimir energy is not the ground state for the new Hamiltonian because,
when $y$ is dynamical, energy can be further lowered by decreasing $y$. 
So to describe Casimir force by a Hamiltonian, 
Casimir energy cannot be a ground-state energy for that Hamiltonian.

\subsection{What has Casimir effect to do with vacuum energy?}

Many physical systems can be approximated by a series of harmonic oscillators. For such systems 
the quantum Hamiltonian takes the form
\begin{equation}\label{ideasH}
 H=\sum_k \hbar\omega_k \left( a_k^{\dagger}a_k+\frac{1}{2} \right) ,
\end{equation}
where $a_k^{\dagger}$ and $a_k$ are raising and lowering operators satisfying $[a_k,a_{k'}^{\dagger}]=\delta_{kk'}$. 
Defining the vacuum $|0\rangle$ as the state annihilated by the lowering operators, $a_k|0\rangle=0$, 
we see that vacuum is also the ground state for the Hamiltonian (\ref{ideasH}).
The corresponding vacuum energy is
\begin{equation}\label{ideasEvac}
 E_{\rm vac}=\langle 0|H|0\rangle = \sum_k \frac{\hbar\omega_k}{2} .
\end{equation}

Now assume that, for some reason, $\omega_k$ depend on some variable $y$. And assume that $y$ is a 
dynamical variable, i.e. the canonical position of some Hamiltonian (say $H+p_y^2/2M$) 
that depends on the canonical momentum $p_y$. Furthermore, assume that all the dependence on $y$ comes from (\ref{ideasH}). 
If these assumptions are fulfilled, then the classical Hamilton equation of motion is
\begin{equation}
 \frac{dp_y}{dt}=-\frac{\partial H}{\partial y} ,
\end{equation}
so in the quantum case we can calculate the average force as 
\begin{equation}\label{ideasFq}
 F=\langle \psi | \left( -\frac{\partial H}{\partial y} \right) |\psi\rangle .
\end{equation}
If $y$ can be treated as a macroscopic classical variable, then (\ref{ideasFq})
can be approximated by
\begin{equation}\label{ideasF}
F = -\frac{\partial}{\partial y} \langle \psi |H|\psi\rangle . 
\end{equation}
Finally if $|\psi\rangle = |0\rangle$, then (\ref{ideasF}) and (\ref{ideasEvac}) give
\begin{equation}\label{ideasF2}
 F=-E'_{\rm vac}(y)=-\sum_k \frac{\hbar\omega_k'(y)}{2} ,
\end{equation}
where the prime denotes derivative with respect to $y$.

Casimir effect can be thought of as an application of (\ref{ideasF2}) to the case where $y$ is
the distance between electrically neutral plates.  

\subsection{Where does the dependence on $y$ come from?}

In the original analysis \cite{casimir} Casimir considered an idealized situation in which 
the plates are made of a perfect conductor with infinite conductivity.
(Such an idealized situation is often the only situation considered in textbooks
\cite{BD,zin-just,mukhanov,zee,schwartz}.)
The electric field vanishes inside the perfect conductor, so electric field between the plates 
must satisfy a boundary condition that enforces the field to vanish at the plates.
Hence the Fourier expansion of the field does not contain wave vectors ${\bf k}$ for which
$k_y$ does not satisfy $k_y=n\pi/y$ (for $n=1,2,3,\ldots)$.
Therefore the frequencies $\omega_{\bf k}$ with $k_y \neq n\pi/y$ do not contribute to 
(\ref{ideasH}) and (\ref{ideasEvac}). 
(By Maxwell equations \cite{jackson} the electric field alone does not oscillate, 
so the existence of frequency really means that we deal with the electromagnetic field.) 
For reasons which will become clear soon, 
instead of saying that those frequencies do not contribute, it is much more appropriate
to say that those frequencies are zero. Therefore we can write
\begin{equation}\label{ideas_omegak}
\omega_{\bf k}=\left\{
\begin{array}{cl}
 c|\bf k| & {\rm for} \;\; k_y=n\pi/y , \\
 0         & {\rm for} \;\; k_y \neq n\pi/y ,
\end{array}
\right.
\end{equation}
where $c$ is the velocity of light. In this way we see that $\omega_{\bf k}$ depend on $y$. 

Of course, realistic materials are usually not perfect conductors. In materials with finite 
conductivity the electric field does not vanish, so boundary conditions do not remove 
wave vectors with $k_y \neq n\pi/y$. A realistic material can be described by a finite dielectric 
constant $\epsilon$. The frequency depends on $\epsilon$ 
as $\omega_{\bf k}=c|\bf k|/\sqrt{\epsilon}$ \cite{jackson}. 
In the vacuum between the plates we have $\epsilon=1$, 
while inside the plates we have $\epsilon\neq 1$. Hence $\epsilon$ is really a function of position 
which, when Fourier transformed, becomes a function of ${\bf k}$ parameterized by the distance $y$. 
In this way, instead of (\ref{ideas_omegak}) we have 
\begin{equation}\label{ideas_omegak2}
 \omega_{\bf k}=\frac{c|\bf k|}{\sqrt{\epsilon_{\bf k}(y)}} .
\end{equation}
That is where the dependence on $y$ comes from in the case of realistic materials.

For perfect conductors the dielectric constant is infinite, so $\epsilon_{\bf k}(y)\rightarrow\infty$
for some values of ${\bf k}$. In this way (\ref{ideas_omegak2})
contains (\ref{ideas_omegak}) as a special case. If we simply ignored wave vectors 
with $k_y \neq n\pi/y$ in (\ref{ideas_omegak}), then we could not see the relation with the
realistic case (\ref{ideas_omegak2}). 

\subsection{Where does $\epsilon$ come from?}

We have seen that Casimir force is related to the dielectric constant $\epsilon$, which 
for perfect conductors is infinite. But dielectric constant is a 
{\em phenomenological macroscopic} quantity, so any description of Casimir effect based on $\epsilon$
lacks the fundamental microscopic origin of the effect.  
A {\em fundamental microscopic} explanation of Casimir effect must involve a microscopic 
explanation of $\epsilon$. Even for perfect conductors, where Casimir effect
can be explained by vanishing electric field in the conductor, one needs to understand 
the {\em microscopic mechanism} by which the field vanishes.

To understand the microscopic origin of $\epsilon$, it is crucial to have in mind that 
electrically neutral materials are made of particles which are {\em electrically charged}.
When electric field is applied to the material, the charges within the material 
rearrange their positions. As a consequence, the local charge density $\rho({\bf x})$ is no longer zero everywhere,
despite that fact that the total charge $\int d^3x\,\rho({\bf x})$ vanishes.
This means that the electric field induces polarization ${\bf P}({\bf x})$ -- 
the electric dipole moment per volume. The polarization itself creates additional electric field, so the 
equations that govern the full electric 
field ${\bf E}({\bf x})$ and polarization ${\bf P}({\bf x})$ become somewhat complicated. 
It turns out \cite{jackson} that it is simpler to describe the system in terms 
of the so-called electric displacement field ${\bf D}({\bf x})$ defined as
\begin{equation}\label{mainD}
 {\bf D}={\bf E}+{\bf P} .
\end{equation}
(For the sake of notational simplicity, we use units in which permittivity of the vacuum is $\epsilon_0=1$.) 
Since ${\bf P}$ is induced by the electric field, it is often a good approximation that 
${\bf P}$ is proportional to ${\bf E}$ \cite{jackson}. Consequently, ${\bf D}$ in (\ref{mainD}) is also
proportional to ${\bf E}$. The dielectric constant $\epsilon$ is defined as that constant of proportionality, 
through the relation
${\bf D}=\epsilon {\bf E}$. Hence (\ref{mainD}) can also be written as
\begin{equation}\label{mainP}
 {\bf P}=(\epsilon -1) {\bf E}.
\end{equation}

\subsection{Where do van der Waals forces come from?}

In a dielectric medium, the energy density associated with electric field is \cite{jackson}
\begin{equation}
 {\cal H}=\frac{{\bf D}\cdot{\bf E}}{2} .
\end{equation}
Using (\ref{mainD}), this can be written as
\begin{equation}\label{main_calH}
 {\cal H}=\frac{{\bf E}^2}{2} + \frac{{\bf P}\cdot{\bf E}}{2} .
\end{equation}
The second term shows that part of energy comes from interaction of electric field 
with {\em polarized charges}. In the absence of an external electric field,
the average field $\langle\psi|{\bf E}|\psi\rangle \equiv \langle{\bf E}\rangle$ vanishes. 
Eq.~(\ref{mainP}) then implies
that $\langle{\bf P}\rangle$ also vanishes, so we have
\begin{equation}\label{main_EP0}
\langle{\bf E}\rangle = \langle{\bf P}\rangle =0 .
\end{equation}
However, unless $|\psi\rangle$ is an electric-field eigenstate,
there are quantum fluctuations $\langle{\bf E}^2\rangle \neq 0$. 
Hence (\ref{mainP}) implies that average interaction energy in (\ref{main_calH}) 
does not vanish
\begin{equation}\label{main_calHint}
\langle{\cal H}_{\rm int}\rangle=\frac{ \langle{\bf P}\cdot{\bf E}\rangle}{2}
=\frac{ \langle{\bf P}^2\rangle }{2(\epsilon -1)}  = \frac{\epsilon -1}{2} \langle{\bf E}^2\rangle .
\end{equation}
This shows that interaction energy, which really originates from the
correlation $\langle{\bf P}\cdot{\bf E}\rangle$ between ${\bf P}$ and ${\bf E}$, can also be related  
to polarization fluctuations $\langle{\bf P}^2\rangle$, or alternatively, to electric field fluctuations 
$\langle{\bf E}^2\rangle$.
Note, however, that the description in terms of fluctuations of ${\bf P}$ or ${\bf E}$
involves the phenomenological macroscopic quantity $\epsilon$. The description in terms of correlations between
${\bf P}$ and ${\bf E}$ is therefore more fundamental because it does not refer to $\epsilon$. 
  
Forces which originate from interaction between correlated electric fields and polarized charges 
obeying (\ref{main_EP0}) are known as van der Waals forces \cite{cohen-tann2,parsegian}. 
In this sense, the interaction energy (\ref{main_calHint}) is nothing but energy of 
van der Waals forces. 

Casimir effect can be described as a force that originates from van der Waals forces.
It is usually described by Lifshitz theory 
\cite{lifshitz,lifshitz2,LL9,milonni,milton,mostepanenko-advances,buhmann1,buhmann2}
which is technically more involved than calculation based on vacuum energy, but eventually 
leads to the same results. In a simple toy model which we shall study,
a Lifshitz-like calculation of the force will turn out to be no more complicated 
than the Casimir-like calculation based on vacuum energy.

\subsection{So what do we need to model Casimir effect from a microscopic perspective?}
\label{SECneed_model}

Now we see that a microscopic description of Casimir effect requires at least 3 dynamical ingredients. 
First, we need the electromagnetic field ${\bf E}({\bf x},t)$ and ${\bf B}({\bf x},t)$. 
Second, we need the polarization field ${\bf P}({\bf x},t)$ originating from microscopic 
charge density $\rho({\bf x},t)$. The electromagnetic and polarization fields are microscopic. 
Third, we need one macroscopic ingredient, namely the distance $y$ between the plates treated
as a dynamical variable.

However, as we said in the Introduction, in this paper we do not want to deal with all the technical details
related to the true Casimir effect. Instead, we want to understand how {\em in general} 
a force can emerge from something which looks like vacuum energy. In other words, we want to understand
how all the main ideas discussed in the present section, Sec.~\ref{SECmain}, are realized in a much simpler
model. For that purpose we shall introduce a single degree of freedom $x_1(t)$ which will mimic
the electromagnetic degrees ${\bf E}({\bf x},t)$ and ${\bf B}({\bf x},t)$. Similarly, another 
single degree of freedom $x_2(t)$ will mimic the polarization field ${\bf P}({\bf x},t)$.
We shall also have the third degree of freedom $y(t)$ which will mimic the distance 
between the plates (even though our model will not describe plates as such). The model 
will be chosen such that, under appropriate approximations, $x_1(t)$ and $x_2(t)$
oscillate with a frequency that depends on $y$.   

Even though the electromagnetic field is different in character from the polarization field, 
we shall take a model which is symmetric under the exchange $x_1\leftrightarrow x_2$. 
It is possible to consider a similar model without such a symmetry, but we enforce this symmetry 
because it greatly simplifies the calculations.
Similarly, to make the calculations as simple as possible, the dependence on $y$ 
will be introduced in a somewhat {\it ad hoc} way. It is possible to introduce a more natural dependence on 
$y$ as in \cite{jaffe2}, but the price is a one-dimensional field theory which has an infinite number of degrees 
of freedom and a divergence that needs to be regularized. 
By contrast, our simple model with only 3 degrees of freedom will not contain any divergence.

\section{The model and its classical properties}
\label{SECmodel}

\subsection{Basic properties of the model}

We consider a system with 3 degrees of freedom $x_1(t)$, $x_2(t)$ and $y(t)$ described by the Hamiltonian
\begin{equation}\label{H}
 H= \left( \frac{p_1^2}{2m}+\frac{kx_1^2}{2} \right) + \left( \frac{p_2^2}{2m}+\frac{kx_2^2}{2} \right)
+ \frac{p_y^2}{2M} + g(y)x_1x_2 .
\end{equation}
Here $p_1$, $p_2$ and $p_y$ are canonical momenta, while $m$, $M$ and $k$ are positive constants. 
(The degrees $x_1$, $x_2$ and $y$ can be interpreted as positions of 3 particles moving in one dimension,
in which case $m$ and $M$ can be interpreted as particle masses, 
but our mathematical results will not depend on that interpretation.)
The function $g(y)$ is an arbitrary non-negative function, 
restricted only by the requirement
\begin{equation}\label{g}
g(y)< k ,
\end{equation}
which provides that energy is positive. 
Our main interest will be the force on the $y$-degree given by
\begin{equation}\label{F}
F=-\frac{\partial H}{\partial y}= -g'(y)x_1x_2 ,
\end{equation}
where the prime denotes derivative with respect to $y$.

\subsection{Normal modes}

For a general $g(y)$, the system cannot be solved exactly in a closed analytic form. Therefore we make 
an approximation. We assume that $y(t)$ changes much slower than $x_1(t)$ and $x_2(t)$, which can be justified 
by taking $M\gg m$. Therefore the motion of $x_1(t)$ and $x_2(t)$ can be found by the adiabatic approximation,
in which their equations of motion are solved by treating $g(y)$ as a constant. This leads to the equations of motion
\begin{eqnarray}
& m\ddot{x}_1(t)+kx_1(t)+gx_2(t)=0 , &
\nonumber \\
& m\ddot{x}_2(t)+kx_2(t)+gx_1(t)=0 . &
\end{eqnarray}
Defining the quantities
\begin{equation}\label{omega}
 \omega^2=\frac{k}{m}, \;\;\; \omega_g^2=\frac{g}{m} ,
\end{equation}
the equations of motion can be written as
\begin{eqnarray}
& \ddot{x}_1(t)+\omega^2 x_1(t)+\omega_g^2 x_2(t)=0 , &
\nonumber \\
& \ddot{x}_2(t)+\omega^2 x_2(t)+\omega_g^2 x_1(t)=0 . &
\end{eqnarray}
This is a system of two coupled oscillators. It can be solved by the ansatz 
\begin{equation}\label{x12}
x_1(t)=c_1e^{-i\Omega t}, \;\;\; x_2(t)=c_2e^{-i\Omega t},
\end{equation} 
which leads to
\begin{eqnarray}\label{c}
& -\Omega^2 c_1 +\omega^2 c_1+\omega_g^2 c_2=0 , &
\nonumber \\
& -\Omega^2 c_2 +\omega^2 c_2 +\omega_g^2 c_1=0 . &
\end{eqnarray}
The first and second equation in (\ref{c}) give
\begin{equation}\label{c2}
 c_2=\frac{\omega_g^2}{\Omega^2-\omega^2}c_1 , \;\;\; c_2=\frac{\Omega^2-\omega^2}{\omega_g^2}c_1 ,
\end{equation}
respectively. These two equations must be consistent, so the factors in front of $c_1$ must be the same. 
This implies $(\Omega^2-\omega^2)^2=\omega_g^4$, i.e. $\Omega^2-\omega^2=\pm \omega_g^2$.
Hence we have two possibilities; either $\Omega^2=\Omega_+^2$ or $\Omega^2=\Omega_-^2$, where
\begin{equation}\label{Omega}
 \Omega_{\pm}^2=\omega^2 \pm\omega_g^2=\frac{k\pm g}{m} ,
\end{equation}
and (\ref{omega}) has been used. We see that $\Omega_{\pm}^2>0$ due to (\ref{g}).
Now both equations in (\ref{c2}) give the same result $c_2=\pm c_1$. Hence we have two 
normal modes of oscillation, namely $x_+(t)$ for which $x_2(t)=x_1(t)$, and $x_-(t)$ for which $x_2(t)=-x_1(t)$.
The most general solution is a superposition of normal modes, so in general we have
\begin{eqnarray}\label{x12pm}
& x_1(t)=c_+x_+(t)+c_-x_-(t) , &
\nonumber \\
& x_2(t)=c_+x_+(t)-c_-x_-(t), &
\end{eqnarray}
where $c_+$ and $c_-$ are arbitrary constants. From (\ref{x12pm})
we see that $x_1(t)$ and $x_2(t)$ are not independent; if we know one 
of them, then we also know the other. Nevertheless, we still have two independent $x$-degrees of freedom, namely 
the normal modes $x_+(t)$ and $x_-(t)$ oscillating with frequencies (\ref{Omega}).

\subsection{Hamiltonian and force in a diagonal form}

Eq.~(\ref{x12pm}) can be used to diagonalize the Hamiltonian (\ref{H}), i.e. to eliminate the term
proportional to $x_1x_2$. By choosing $c_+=c_-=1/\sqrt{2}$ in (\ref{x12pm}) we have 
\begin{equation}\label{x12diag}
 x_1=\frac{x_+ + x_-}{\sqrt{2}} , \;\;\; x_2=\frac{x_+ - x_-}{\sqrt{2}} ,
\end{equation}
the inverse of which is
\begin{equation}\label{xpm}
 x_{\pm}=\frac{x_1\pm x_2}{\sqrt{2}} .
\end{equation}
We can think of (\ref{x12diag}) and (\ref{xpm}) not as solutions to the equations of motion, but as a definition 
of the new canonical coordinates $x_+$ and $x_-$.
We see that
\begin{equation}
 \frac{k(x_1^2+x_2^2)}{2}+g(y)x_1x_2=\frac{k_+(y)x_+^2}{2}+\frac{k_-(y)x_-^2}{2} ,
\end{equation}
where 
\begin{equation}
 k_{\pm}(y)=k\pm g(y) .
\end{equation}
Hence (\ref{H}) can be written as
\begin{equation}\label{H2}
 H=H_+ + H_- +\frac{p_y^2}{2M} ,
\end{equation}
where
 \begin{equation}\label{Hpm}
 H_{\pm}= \frac{p_{\pm}^2}{2m}+\frac{k_{\pm}(y)x_{\pm}^2}{2} .
\end{equation}
If we neglect the last term $p_y^2/2M$ in (\ref{H2}), i.e. treat $y$ as a non-dynamical constant, we see immediately 
that (\ref{Hpm}) leads to the oscillatory solutions with frequencies (\ref{Omega}).
In addition, using (\ref{xpm}) one can see that
\begin{equation}\label{ppm}
 p_{\pm}=\frac{p_1\pm p_2}{\sqrt{2}} .
\end{equation}

Now let us discuss the force (\ref{F}). Inserting (\ref{x12diag}) into (\ref{F}) we get
\begin{equation}\label{F2}
F= -\frac{g'(y)\, (x_+^2-x_-^2)}{2} .
\end{equation}
We are interested in the force when $x_1$ and $x_2$ are in their ground state. From (\ref{Hpm}) we see that they are
in the classical ground state when $x_{\pm}=0$. Hence (\ref{F2}) implies that $F=0$ when 
$x_1$ and $x_2$ are in their classical ground state.

\section{Quantum force}
\label{SECforce}

The full interacting Hamiltonian (\ref{H}) has a quantum ground state with some finite ground-state energy $E_{\rm vac}$.
This energy is a number that does not depend on $x_1$, $x_2$ and $y$. As such, it does not have any 
physical consequences so can be subtracted from the Hamiltonian without affecting physics.
The ground state of the full Hamiltonian (\ref{H}) does not contain any interesting physics.

However, interesting physics may appear when the system is in an {\em effective} ground state, in which
{\em some but not all} degrees of freedom are in their ground state.
As in the discussion after (\ref{F2}),
we shall be interested in the case when $y$ is {\em not} in the ground state, while $x_1$ and $x_2$ are.    

\subsection{Quantization of the free Hamiltonian}
\label{Qfree}

As a warm up, let us quantize (\ref{H}) in the free case $g(y)=0$. 
In this case we have two uncoupled
harmonic oscillators with frequency $\omega$ given in (\ref{omega}), which is a well-known textbook stuff
(see e.g. \cite{cohen-tann1}). One introduces the operators 
\begin{equation}\label{a}
 a_j=\sqrt{\frac{m\omega}{2\hbar}}x_j+\frac{i}{\sqrt{2m\hbar\omega}}p_j ,
\end{equation}
for $j=1,2$, which satisfy $[a_j,a^{\dagger}_{j'}]=\delta_{jj'}$. The position and momentum operators can be expressed from
(\ref{a}) as
\begin{equation}\label{xp}
 x_j=\sqrt{\frac{\hbar}{2m\omega}}(a^{\dagger}_j+a_j) , \;\;\;
p_j=i\sqrt{\frac{m\hbar\omega}{2}}(a^{\dagger}_j-a_j) , 
\end{equation}
so (\ref{H}) with $g(y)=0$ can be written as
\begin{equation}\label{Hfree}
 H^{\rm (free)}=H_1+H_2+\frac{p_y^2}{2M} ,
\end{equation}
where
\begin{equation}\label{Hj}
H_j=\hbar\omega\left( a^{\dagger}_ja_j+\frac{1}{2} \right).
\end{equation}

The last term in (\ref{Hfree}) can also be quantized, but we shall use an approximation in which
$y$ is treated as a classical variable. This can be justified by assuming that $y$ is a macroscopic
degree of freedom, with a large mass $M\gg m$. This means that we are really doing a semi-classical theory,
in which $y(t)$ is treated as a classical background. 
Hence we only quantize the Hamiltonian 
\begin{equation}\label{Heff}
H^{\rm (free\; eff)}=H_1+H_2 ,
\end{equation}
which is the free effective Hamiltonian for $x_1$ and $x_2$ degrees.
The corresponding free effective vacuum $|0\rangle$ satisfies
\begin{equation}\label{0}
 a_j|0\rangle=0,
\end{equation}
so the free effective-vacuum energy is
\begin{equation}
 E^{\rm (free\; eff)}_{\rm vac}=\langle 0|H^{\rm (free\; eff)}|0\rangle =
\frac{\hbar\omega}{2}+\frac{\hbar\omega}{2} .
\end{equation}
Clearly this free effective-vacuum energy is a constant which does not depend on $y$. As such, it does not have any 
physical consequences so can be subtracted from the Hamiltonian without affecting physics.

\subsection{Force {\it \`a la} Casimir}

In the interacting case, the full Hamiltonian is (\ref{H2}). As in the free case in Sec.~\ref{Qfree},
we treat $y$ as a classical background. Therefore we quantize only the effective Hamiltonian
\begin{equation}\label{Heff2}
H^{\rm (eff)}=H_+ +H_- ,
\end{equation}
which is the interacting version of (\ref{Heff}). Analogously to (\ref{a}) and (\ref{xp}) we have
\begin{equation}\label{a2}
 a_{\pm}=\sqrt{\frac{m\Omega_{\pm}}{2\hbar}}x_{\pm}+\frac{i}{\sqrt{2m\hbar\Omega_{\pm}}}p_{\pm} ,
\end{equation}
\begin{equation}\label{xp2}
 x_{\pm}=\sqrt{\frac{\hbar}{2m\Omega_{\pm}}}(a^{\dagger}_{\pm}+a_{\pm}) , \;\;\;
p_{\pm}=i\sqrt{\frac{m\hbar\Omega_{\pm}}{2}}(a^{\dagger}_{\pm}-a_{\pm}) , 
\end{equation}
so (\ref{Hpm}) gives
\begin{equation}\label{Hpm2}
H_{\pm}=\hbar\Omega_{\pm}\left( a^{\dagger}_{\pm}a_{\pm}+\frac{1}{2} \right),
\end{equation}
which is the interacting version of (\ref{Hj}).
Analogously to (\ref{0}), the interacting effective vacuum $|\tilde{0}\rangle$ satisfies
\begin{equation}\label{t0}
 a_{\pm}|\tilde{0}\rangle=0 ,
\end{equation}
so the interacting effective-vacuum energy is
\begin{equation}\label{Evac}
 E^{\rm (eff)}_{\rm vac}=\langle \tilde{0}|H^{\rm (eff)}|\tilde{0}\rangle =
\frac{\hbar\Omega_+(y)}{2}+\frac{\hbar\Omega_-(y)}{2} .
\end{equation}

Note that $\Omega_{\pm}(y)$ depend on $y$ because $\Omega_{\pm}$ depend on $g$ due to (\ref{Omega}),
and $g$ depends on $y$ as we assumed already in (\ref{H}).
Hence the force in the interacting effective vacuum can be calculated as
\begin{equation}\label{Fcas}
 F=-\frac{\partial E^{\rm (eff)}_{\rm vac}}{\partial y}= -\frac{\hbar\Omega'_+(y)}{2}-\frac{\hbar\Omega'_-(y)}{2} .
\end{equation}
From (\ref{Omega}) we see that
\begin{equation}
 \Omega'_{\pm}=\frac{\pm g'}{2m\Omega_{\pm}} ,
\end{equation}
so (\ref{Fcas}) becomes
\begin{equation}\label{Fcas2}
 F=\frac{-\hbar g'(y)}{4m\Omega_+(y)}+\frac{\hbar g'(y)}{4m\Omega_-(y)} .
\end{equation}

\subsection{Force {\it \`a la} Lifshitz}

One may be worried that the calculation of the force based on (\ref{Fcas})
looks somewhat {\it ad hoc} \cite{nik_casimir}, because it is not clear 
how the force (\ref{Fcas}) is related to the canonical way to calculate the force
by (\ref{F}) or (\ref{F2}). It is more legitimate to calculate the quantum force as
the expectation value of the force operator, namely
\begin{equation}\label{Flifs}
 F=-\langle\Psi|g'(y)x_1x_2 |\Psi\rangle = -\frac{ \langle\Psi|g'(y)\, (x_+^2-x_-^2)|\Psi\rangle}{2} ,
\end{equation}
where $|\Psi\rangle$ is the full quantum state of the system. We are using the approximation in which 
$y$ is treated as a classical background while $x_1$ and $x_2$ are in the interacting effective vacuum
$|\tilde{0}\rangle$, so (\ref{Flifs}) can be approximated by
\begin{equation}\label{Flifs2}
 F=-\frac{ g'(y)\, \langle\tilde{0}| (x_+^2-x_-^2)|\tilde{0}\rangle}{2} .
\end{equation}
From the first equation in (\ref{xp2}) we see that
\begin{equation}\label{p2}
\langle\tilde{0}|x_{\pm}^2 |\tilde{0}\rangle = \frac{\hbar}{2m\Omega_{\pm}} ,
\end{equation}
so (\ref{Flifs2}) becomes 
\begin{equation}\label{Flifs3}
 F=\frac{-\hbar g'(y)}{4m\Omega_+(y)}+\frac{\hbar g'(y)}{4m\Omega_-(y)} .
\end{equation}
We see that this result coincides with (\ref{Fcas2}).

Some remarks are in order. First, in the calculation of (\ref{Flifs3}) we never referred 
to the energy of the vacuum. We did, however, referred to the vacuum value of $x_{\pm}^2$
in (\ref{p2}). If we used the normal ordered product $:\! x_{\pm}^2 \!:$ instead of $x_{\pm}^2$
we would get a zero force, which would be a wrong result. The quantum fluctuations 
described by (\ref{p2}) are physical. 

Second, Eq.~(\ref{Flifs}) makes it clear that the force originates from the interaction between 
$x_1$, $x_2$, and $y$. The fact that interaction between $x_1$ and $x_2$ is important 
is not so clear from the calculation based on (\ref{Fcas}). In this sense, even though both 
calculations lead to the same result, the calculation based on (\ref{Flifs})
better reflects the true physical origin of the force. 

Third, note that in the free vacuum $|0\rangle$ we have 
\begin{equation}
 \frac{\langle 0| (x_+^2-x_-^2)|0\rangle}{2}=\langle 0| x_1x_2|0\rangle=0. 
\end{equation}
This means that $x_1$ and $x_2$ are not correlated in the free vacuum, which is why 
the force vanishes in the free vacuum. More generally, if we considered a state of the form
$|\psi\rangle = |0_1\rangle |\psi_2\rangle$, so that only $x_1$ is in the free vacuum 
while $x_2$ is in an arbitrary quantum state, we would again get
\begin{equation}\label{x1x2cor}
 \langle \psi| x_1x_2|\psi\rangle=0 , 
\end{equation}
so the force would vanish again. To get any force at all, it is important that $x_1$
is {\em not} in the free vacuum state. Similar, of course, is also true for $x_2$.

\section{The content of the interacting vacuum}
\label{SECcontent}
  
\subsection{General remarks}

Any quantum state of $x_1$ and $x_2$, with or without interaction, can be expanded in the complete basis
$|n,n'\rangle$ defined by
\begin{equation}
 |n,n'\rangle = \frac{(a_1^{\dagger})^n (a_2^{\dagger})^{n'} |0\rangle}{\sqrt{n!n'!}} .
\end{equation}
The interacting vacuum $|\tilde{0}\rangle$ is not an exception, so there are some coefficients $c_{nn'}$ 
such that
\begin{equation}\label{tilde0}
 |\tilde{0}\rangle=\sum_{n=0}^{\infty} \sum_{n'=0}^{\infty} c_{nn'}|n,n'\rangle .
\end{equation}

\subsection{Bogoliubov transformation}

The coefficients $c_{nn'}$ can be found by noting that the operators $a_{\pm}$ and $a_{\pm}^{\dagger}$
are related to $a_j$ and $a_j^{\dagger}$ by a Bogoliubov transformation. 
To see that, we insert (\ref{xpm}) and (\ref{ppm}) into (\ref{a2}) and use (\ref{xp}) to obtain
\begin{equation}\label{Bogol}
a_{\pm}=\sum_{j=1,2}\alpha_{j\pm}a_1 + \beta_{j\pm}a_j^{\dagger} ,
\end{equation}
where
\begin{eqnarray}\label{alphabeta}
& \alpha_{1\pm}=\displaystyle\frac{\Omega_{\pm}+\omega}{2\sqrt{2\Omega_{\pm}\omega}} ,
\;\;\; \alpha_{2\pm}=\pm\alpha_{1\pm} , &
\nonumber \\
& \beta_{1\pm}=\displaystyle\frac{\Omega_{\pm}-\omega}{2\sqrt{2\Omega_{\pm}\omega}} ,
\;\;\; \beta_{2\pm}=\pm\beta_{1\pm} , &
\end{eqnarray}
are the Bogoliubov coefficients. Bogoliubov coefficients, in general, can be complex,
but in our case they are real. They satisfy
\begin{equation}
 \sum_{j=1,2}(\alpha_{j\pm}^2 - \beta_{j\pm}^2)=1 .
\end{equation}
The inverse transformation of (\ref{Bogol}) is 
\begin{equation}\label{Bogol2}
a_j=\sum_{s=+,-}\alpha_{js}a_s - \beta_{js}a_s^{\dagger} .
\end{equation}
This implies $a_j^{\dagger}=\sum_{s}\alpha_{js}a_s^{\dagger} - \beta_{js}a_s$, so one finds
\begin{equation}\label{N}
\langle\tilde{0}|a_j^{\dagger}a_j |\tilde{0}\rangle = \sum_{s=+,-}\beta_{js}^2 .
\end{equation}
The number of free quanta is counted by the operators $N_j=a_j^{\dagger}a_j$, 
so (\ref{N}) tells us that the average number of free quanta in $|\tilde{0}\rangle$ is non-zero
\begin{equation}\label{N2}
 \langle\tilde{0}|N_j |\tilde{0}\rangle = \beta_{j+}^2 +\beta_{j-}^2 .
\end{equation}

\subsection{The expansion coefficients}

In our case $\beta_{j+}$ and $\beta_{j-}$ are different and both non-zero. 
This makes the explicit calculation of $c_{nn'}$ in (\ref{tilde0}) somewhat complicated.  
Physically, this is related to the fact that both $\Omega_+$ and $\Omega_-$
contribute to (\ref{Evac}). As we shall see in Sec.~\ref{SECreal}, 
in the real Casimir effect there are reasons to ignore the contribution from one frequency (say $\Omega_-$)
and consider only the contribution from the other (say $\Omega_+$). From this perspective it makes sense
to ignore $\beta_{j-}$ and $\alpha_{j-}$ and to consider only 
$\beta_{j+}\equiv\beta$ and $\alpha_{j+}\equiv\alpha$. Hence, instead of (\ref{Bogol}) 
we study a simplified Bogoliubov transformation
\begin{equation}\label{Bogols}
a=\alpha(a_1+a_2) + \beta(a_1^{\dagger}+a_2^{\dagger}).
\end{equation}
Such a Bogoliubov transformation is known to lead to the so-called two-mode squeezed states \cite{schumaker},
but here we shall study it from scratch without referring to the results in the literature.

The interacting vacuum $|\tilde{0}\rangle$ is defined by
\begin{equation}\label{a0}
 a|\tilde{0}\rangle=0 .
\end{equation}
Instead of starting from the general expansion (\ref{tilde0}), we make the ansatz 
\begin{equation}\label{tilde02}
 |\tilde{0}\rangle=\sum_{n=0}^{\infty} c_{n}|n,n\rangle .
\end{equation}
By inserting (\ref{Bogols}) and (\ref{tilde02}) into (\ref{a0}) and using
\begin{eqnarray}
& a_1|n,n\rangle = \sqrt{n}|n-1,n\rangle , &
\nonumber \\
& a_2|n,n\rangle = \sqrt{n}|n,n-1\rangle , &
\nonumber \\
& a_1^{\dagger}|n,n\rangle = \sqrt{n+1}|n+1,n\rangle , &
\nonumber \\
& a_2^{\dagger}|n,n\rangle = \sqrt{n+1}|n,n+1\rangle , &
\end{eqnarray}
we obtain 
\begin{eqnarray}
& \displaystyle\sum_{n=1}^{\infty}\alpha c_n \sqrt{n} (|n-1,n\rangle+|n,n-1\rangle) &
\nonumber \\
& + \displaystyle\sum_{n=0}^{\infty}\beta c_n \sqrt{n+1} (|n+1,n\rangle+|n,n+1\rangle) =0.& 
\end{eqnarray}
In the first sum we introduce a new variable $n'=n-1$ and then remove the prime from $n'$ 
because it is a dummy variable. This leads to
\begin{equation}
 \displaystyle\sum_{n=0}^{\infty} [\alpha c_{n+1} + \beta c_n] \, \sqrt{n+1}\,  (|n+1,n\rangle+|n,n+1\rangle) =0. 
\end{equation}
Hence the expression in the square brackets must vanish, which leads to the simple recursion relation 
$c_{n+1}=-(\beta/\alpha)c_n$ with the solution
\begin{equation}
 c_n=\left(-\frac{\beta}{\alpha×}\right)^n c_0. 
\end{equation}
Therefore (\ref{tilde02}) becomes
\begin{equation}\label{tilde03}
 |\tilde{0}\rangle=c_0 \sum_{n=0}^{\infty} \left(-\frac{\beta}{\alpha}\right)^n |n,n\rangle .
\end{equation}
The constant $c_0$ can be determined from the normalization condition $\langle \tilde{0}|\tilde{0}\rangle=1$. 
This gives
\begin{equation}
 |c_0|^2 \sum_{n=0}^{\infty} \left(-\frac{\beta}{\alpha}\right)^{2n} =1 ,
\end{equation}
so applying the geometric series formula $\sum_{n=0}^{\infty} z^n =(1-z)^{-1}$ 
to $z=(-\beta/\alpha)^2=(\beta/\alpha)^2$, we get
\begin{equation}
 c_0=\sqrt{1-(\beta/\alpha)^2} .
\end{equation}

Note that $|n,n\rangle$ in (\ref{tilde03}) is a state in which the number of $x_1$-quanta is always equal to the number
of $x_2$-quanta. In other words, $|n,n\rangle$ describes $n$ pairs, 
where each pair contains one $x_1$-quantum and one $x_2$-quantum. 
The interacting vacuum $|\tilde{0}\rangle$ is a state with an uncertain number of 
such pairs. In the real Casimir effect, to which we turn in the next section,
$x_1$-quantum corresponds to a photon and $x_2$-quantum corresponds to a quantum of polarization. 

\section{Relation to the real Casimir effect}
\label{SECreal}

Let us now discuss how the calculations in our toy model 
are related to the calculations for the real Casimir effect.
Our discussion will be non-technical, qualitative and hopefully intuitive.

As we already said in Sec.~\ref{SECneed_model}, $x_1$ mimics electromagnetic field and 
$x_2$ mimics polarization field. For instance, the first bracket in (\ref{H}) is analogous
to the pure electromagnetic Hamiltonian, i.e. we have the analogy 
\begin{equation}
\frac{p_1^2/m + kx_1^2}{2} \; \leftrightarrow \; \int d^3x\, \frac{{\bf E}^2+{\bf B}^2}{2} .
\end{equation}
Similarly, the second bracket in (\ref{H}) is analogous
to the pure matter term. For an appropriate model of polarized matter in a dielectric material
see e.g. \cite{hopfield}.  
The last term in (\ref{H}) is similar to the interaction between charges and electromagnetic field, i.e.
\begin{equation}\label{anal2}
 gx_1x_2 \; \leftrightarrow \;  \int d^3x\,A_{\mu}j^{\mu} ,
\end{equation}
where $A_{\mu}$ is the electromagnetic 4-potential and $j^{\mu}$ is the charged 4-current.

The interaction between charged matter and electromagnetic field in terms of normal modes 
and their frequencies is discussed in many solid-state textbooks 
\cite{kittel,ashcroft-mermin,callaway,solyom,fox,mahan,grosso}.
Instead of two frequencies $\Omega_{\pm}$ in (\ref{Omega})
one gets two branches of the dispersion relation $\omega_{\pm}({\bf k})$.
The upper branch $\omega_+({\bf k})$ is phonon-like (i.e. varies very slowly with ${\bf k}$)
for small $|{\bf k}|$ and photon-like (i.e. behaves approximately as $\omega_+({\bf k})\approx c|{\bf k}|$)
for large $|{\bf k}|$. The lower branch $\omega_-({\bf k})$ has the opposite behavior, i.e.
it is  photon-like for small $|{\bf k}|$ and phonon-like for large $|{\bf k}|$. 
Thus, ignoring a small range of intermediate $|{\bf k}|$, we can use an approximation according to which,
for each ${\bf k}$,
we have only one branch that significantly varies with ${\bf k}$. Since Casimir force (\ref{ideasF2}) 
is proportional to 
\begin{equation}
\int d^3k \, \frac{\partial\omega({\bf k})}{\partial y} = 
\int d^3k \, \frac{\partial\omega({\bf k})}{\partial {\bf k}} \frac{\partial{\bf k}}{\partial y} ,
\end{equation}
it follows that for each ${\bf k}$ there is only one branch with dispersion relation $\omega({\bf k})\approx c|{\bf k}|$
that significantly contributes to Casimir force. 
Indeed, $\partial\omega({\bf k})/\partial {\bf k}$ is the group velocity of waves, which for phonons is much smaller 
than that for photons. A calculation for perfect conductors (see e.g. \cite{itz-zub}) gives the Casimir force 
\begin{equation}\label{realF}
 F(y) = -\frac{\pi^2}{240}\frac{\hbar c}{y^4} ,
\end{equation}
so heuristically one may expect that a contribution from the phonon-like branch 
would give a similar result with $c\rightarrow c_s$,
which would be negligible because the velocity of sound $c_s$ is much smaller than the velocity of light $c$. 
More details about calculations of Casimir force from vacuum energy of realistic materials can be found in
\cite{schram,spruch,klimchitskaya_etal,milonni}.  

Not let us say a few words about the physical nature of normal modes. Just like 
$x_{\pm}$ in (\ref{xpm}) is a mixture of $x_1$ and $x_2$, a normal mode in a real material is
a mixture of electromagnetic field and polarization field \cite{hopfield}. 
For a photon-like branch, (\ref{xpm}) is roughly analogous to (\ref{mainD}),
as studied in more detail in \cite{philbin1} and applied to Casimir effect in \cite{philbin2}.
As analyzed in \cite{hopfield}, the number of quanta of such mixed fields 
are lowered and raised by operators analogous to (\ref{a2}). Such mixed quanta are often referred to as 
polaritons \cite{hopfield-henry,kittel,ashcroft-mermin,callaway,solyom,fox,mahan,grosso}.
(Note, however, that the word ``polariton'' was first introduced in \cite{hopfield} where 
it meant the quantum of pure polarization field, not the quantum of a mixture.)
The lowering and raising operators for mixed fields are related to lowering and raising operators 
for electromagnetic and polarization fields by a Bogoliubov transformation \cite{hopfield}
analogous to that in Sec.~\ref{SECcontent}. The Casimir vacuum can be expressed in terms of photons and 
polarization quanta by a Bogoliubov transformation \cite{ciccarello,passante}
leading to a state analogous to (\ref{tilde03}). This means that Casimir vacuum can be thought of as a state
with a zero number of polaritons \cite{simpson}, but one should not forget that this vacuum
is really a state with an uncertain number of pairs, 
with each pair containing one photon and one quantum of polarization. 
Nevertheless, these photons and polarization quanta cannot be
directly observed because they are not energy eigenstates of the interacting Hamiltonian.  
The energy eigenstates are polaritons.
For other roles of polaritons in Casimir physics see also 
\cite{lenac-tomas1,lenac-tomas2,apostol,haakh,tercas}.

Finally a few words on Lifshitz theory. Analogously to (\ref{F}), one can start from the
classical Lorentz force on charges derived from electromagnetic interaction (\ref{anal2}).
In the quantum case, analogously to the first equality in (\ref{Flifs}), 
the average Lorentz force is  
\begin{equation}
 {\bf F}=\int d^3x\, \langle \rho{\bf E}+ {\bf j}\times{\bf B} \rangle .
\end{equation}
By Maxwell equations, $\rho$ and ${\bf j}$ can be expressed in terms of ${\bf E}$ and ${\bf B}$.
In this way the calculation of force reduces to a calculation of the expectation value 
of an operator quadratic in ${\bf E}$ and ${\bf B}$. This operator
turns out to be proportional to a derivative of the energy-momentum tensor of electromagnetic field 
\cite{buhmann1,buhmann2}. The explicit calculation is technically involved 
\cite{lifshitz,lifshitz2,LL9,milonni,milton,mostepanenko-advances,buhmann1,buhmann2},
but is conceptually analogous to the simple calculation (\ref{Flifs})-(\ref{Flifs3}).
The result (\ref{x1x2cor}) is analogous to the result \cite{nik_casimir} 
that $\langle A_{\mu}j^{\mu} \rangle$ vanishes in any state proportional to the photon vacuum,
implying that Casimir force is impossible in photon vacuum \cite{nik_casimir}. 

\section{Conclusions}
\label{SECconcl}

There is a general physical principle telling that zero-point energy is unphysical.
Casimir effect is perfectly compatible with this principle.
In this paper we have seen that this can be understood at several levels.

First, Casimir vacuum is not a ground state for the full Hamiltonian, but only
a ground state for an effective Hamiltonian that describes physics at a fixed distance $y$
between Casimir plates. A description of Casimir force requires $y$ to be a dynamical variable,
and Casimir vacuum is not a ground state for a Hamiltonian in which $y$ is dynamical.

Second, at a fundamental microscopic level, Casimir force should be viewed as a manifestation
of van der Waals forces, which involves correlated fluctuations of polarization 
and associated electromagnetic field.  

Third, Casimir vacuum is not a state without photons. It can be related to the photon 
vacuum by a non-trivial Bogoliubov transformation,  
leading to the picture of Casimir vacuum as a state with a zero number of certain
quasi-particle excitations (polaritons), but uncertain number of photons and polarization quanta.  

\section*{Acknowledgments}

The author is grateful to R. Erdem for comments on the manuscript.
This work was supported by the Ministry of Science of the Republic of Croatia
and by H2020 Twinning project No. 692194, ``RBI-T-WINNING''.

\end{document}